\documentclass[12pt]{article}
\usepackage{amsfonts}
\begin{document}

\def\href#1#2{#2}
\newpage
\pagestyle{empty}
\setcounter{page}{0}
\vspace{20mm}

\begin{center}

  {\LARGE  {\sffamily 
      Algebraic approach to  $q$-deformed supersymmetric
        variants \\[.5cm] of the Hubbard model with pair hoppings
      }}\\[1cm]
  
  \vspace{10mm}
  
  {\large Daniel Arnaudon \footnote{Daniel.Arnaudon@lapp.in2p3.fr.}}\\[.42cm]
  
  {\em Laboratoire de Physique Th{\'e}orique }ENSLAPP\footnote{URA 1436 
    du CNRS, associ{\'e}e {\`a} l'{\'E}cole Normale Sup{\'e}rieure de Lyon et
    {\`a} l'Universit{\'e} de Savoie.
    \newline\indent
    Work partially supported by European Community contract TMR
    FMRX-CT96.0012}, CNRS\\[.242cm]
  
  Groupe d'Annecy: LAPP, BP 110, F-74941
  Annecy-le-Vieux Cedex, France.
  \\

\end{center}
\vspace{20mm}

\begin{abstract}
  We construct two quantum spin chains
  Hamiltonians with quantum $sl(2|1)$ invariance. 
  These spin chains define variants of the Hubbard model and describe
  electron models with pair hoppings.  
  A cubic algebra that admits   the Birman--Wenzl--Murakami
  algebra as a quotient allows exact solvability of the periodic chain. 
  The two Hamiltonians, respectively built using the distinguished and
  the fermionic bases of ${\cal U}_q(sl(2|1))$ differ only in the boundary
  terms. They are actually equivalent, but the equivalence is non local.
  Reflection equations are solved to get exact solvability on open
  chains with non trivial boundary conditions. Two families of
  diagonal solutions are found. 
  The centre and the Scasimirs of the quantum enveloping algebra of 
  $sl(2|1)$ appear as tools for the construction of exactly solvable
  Hamiltonians. 
\end{abstract}

\rightline{physics/9711001}
\rightline{ENSLAPP-A-650/97}
\rightline{October 1997}
\rightline{Published in JHEP Electronic Journal: JHEP 12(1997)006}

\newpage
\pagestyle{plain}
%%%%%%%%%%%%%%%%%%%%%%%%%%%%%%%%%%%%%%%%%%%%%%%%%%%%%%%%%%%%%%%%%%%%

%\def\mathbb{\bf}
\def\CC{{\mathbb C}}
\def\NN{{\mathbb N}}
\def\QQ{{\mathbb Q}}
\def\RR{{\mathbb R}}
\def\ZZ{{\mathbb Z}}
\def\cA{{\cal A}}          \def\cB{{\cal B}}          \def\cC{{\cal C}}
\def\cD{{\cal D}}          \def\cE{{\cal E}}          \def\cF{{\cal F}}
\def\cG{{\cal G}}          \def\cH{{\cal H}}          \def\cI{{\cal I}}
\def\cJ{{\cal J}}          \def\cK{{\cal K}}          \def\cL{{\cal L}} 
\def\cM{{\cal M}}          \def\cN{{\cal N}}          \def\cO{{\cal O}}
\def\cP{{\cal P}}          \def\cQ{{\cal Q}}          \def\cR{{\cal R}} 
\def\cS{{\cal S}}          \def\cT{{\cal T}}          \def\cU{{\cal U}}
\def\cV{{\cal V}}          \def\cW{{\cal W}}          \def\cX{{\cal X}}
\def\cY{{\cal Y}}          \def\cZ{{\cal Z}}
%
%%\bibliographystyle{utphys}
%%\bibliographystyle{unsrt}
%\def\href#1#2{#2}
%
% D{\'e}finitions locales:
%
\def\Id{\mbox{Id}}
\def\matdiag#1#2#3#4{ \left(
  \begin{array}{cccc} #1 &&&\\ &#2&&\\ &&#3&\\ &&&#4 \end{array} 
  \right)}
\def\qmbox#1{\qquad\mbox{#1}\qquad}
\def\uq#1{{\cal U}_q(sl(#1))}
\def\Vect{\mbox{Vect}}
\def\vR{\check\cR}
\renewcommand{\thefootnote}{\fnsymbol{footnote}}

%\rightline{physics/9711001}
%\rightline{\ENSLAPP-A-650/97}
%\rightline{October 1997}

\section{\label{sect:introduction}Introduction}

Since a few years, there is a considerable interest about some models
of strongly correlated electrons, in particular those of the families of the
$t$--$J$ model and of the Hubbard model. The reason is that they exhibit
some very interesting physical properties related with 
high $T_c$ superconductivity.
Among these models, some have the property of supersymmetry, or
quantum supersymmetry. This is the case for some generalisations 
of the $t$--$J$ model. It is also the case for some variants  
of the Hubbard models in which a pair hopping term is included
(\cite{EssKorSch:prl68,EssKorSch:prl70,BGLZ:prl74} and 
\cite{GHLZ:pla212} for quantum
supersymmetry). 

The aim of this paper is to present the construction of two variants
of the supersymmetric Hubbard model with pair hoppings, to describe
the algebra that ensures their integrability and to solve the reflection
equations which lead to integrable boundary terms.

{}From the expression of the series of Casimir operators $\cC_p$ of
$\uq{2|1}$, we derive quantum spin chain Hamiltonians $\cH$ with 
built-in $\uq{2|1}$ invariance:
\begin{equation}
  \cH = \sum_{i=1}^{L-1} 1\otimes \cdots \otimes 
  \underbrace{(\rho\otimes\rho)\Delta( \mbox{Pol} \{\cC_p \} )}_{
    \mbox{sites\ } i,i+1}
  \otimes \cdots \otimes 1 \;.
  \label{eq:Hintro}
\end{equation}

An ingredient of the construction is the knowledge of a series of
Casimir operators of the quantum algebra. We will also see that the
knowledge of Scasimirs (given in section \ref{sect:sl12}) 
leads to some exactly solvable Hamiltonians. 

Another ingredient of the construction is the four
dimensional one parameter typical representation $\rho$ of $\uq{2|1}$, so
that the Hamiltonian describes a four states per site spin chain with
two parameters (the parameter of the representation together with the
deformation parameter $q$). 

The integrability of the closed chain is based on the algebra
\begin{eqnarray}
  && \left(b_i + q \right) \left(b_i - q \lambda^2 \right)
  \left(b_i - q^{-1} \lambda^{-2} 
  \right) = 0  \;,  \label{eq:algebra1}
  \\
  && b_i b_{i\pm 1} b_i  =  b_{i\pm 1} b_i b_{i\pm 1} \;, \\
  && b_i b_j  =  b_j b_i \qmbox{for} |i-j|\ge 2 \;, \\
  &&  (b_i-x) b_{i\pm 1}^{-1} (b_i -x)
  - b_i^{-1} (b_{i\pm 1}-x) b_i^{-1} 
  =
  (b_{i\pm 1}-x) b_i^{-1} (b_{i\pm 1} -x)
  - b_{i\pm 1}^{-1} (b_i-x) b_{i\pm 1}^{-1} \;.
  \nonumber\\ &&
  \label{eq:algebra4}
\end{eqnarray}
This algebra was proved in \cite{GeWuXue:1991} to be sufficient to
construct a solution $\vR(u)$ of the Yang--Baxter algebra (see below
(\ref{eq:YBA})). Moreover, the Birman--Wenzl--Murakami algebra
\cite{Birman-Wenzl,Murakami} 
is a quotient of this algebra. Our realisation of the algebra
(\ref{eq:algebra1}--\ref{eq:algebra4}) actually does not satisfy the
supplementary relations of the BWM algebra. 
The operators $b_i$ enters in the expression of the two site Hamiltonian as 
\begin{equation}
  \cH_{i,i+1} = b_i - b_i^{-1} \;.
  \label{eq:H=b-1/b}
\end{equation}

A remarkable fact is that, using the distinguished and fermionic
bases of $\uq{2|1}$, we obtain two different Hamiltonians, the
difference being in the boundary terms.  
The same phenomenon was described in \cite{ACF} with three state per
site spin chains (deformed supersymmetric $t$--$J$ model). 
These Hamiltonians are actually equivalent
{\em on open chains}, but this equivalence, which comes from a
twist of the coproduct, is
non trivial since it is non local on the chain.

One of the Hamiltonians (constructed with the distinguished basis) 
was known to be exactly solvable \cite{BarKluZit,GHLZ:pla212}. 
It was obtained in \cite{GHLZ:pla212}, starting from
the expression of the spectral parameter $\cR$-matrix of $\uq{2|1}$.

The reflection equations associated with the solution $\vR(u)$ of the
Yang--Baxter algebra are solved for diagonal $\cK$ matrices. Two
families of one parameter solutions are found for each equation,
leading to four possible boundary terms for exactly solvable open
chain Hamiltonians. This number of solutions is the same as found in
\cite{Gonzalez-Ruiz:npb424} in the case of the supersymmetric $t$--$J$
model.
It is then shown that a special choice of these
boundary terms is exactly the difference of the two Hamiltonians built
from the distinguished and the fermionic bases.

In the Appendix, the expressions of the Scasimir operators of the (non
quantised) $sl(2|1)$ superalgebra are given.

This work was already completed when the paper \cite{BGZZ:cond-mat9710}
appeared. In this paper, the Hamiltonian (\ref{eq:Hdist})
corresponding to the 
distinguished basis is studied. One of the solutions (i.e. \ref{eq:K-sol1})
for the reflection equations is given and the corresponding integrable
boundary terms are computed. 
The Bethe ansatz equations are also written. 
Analogous results were also obtained in \cite{ZZ:cond-mat9707} for the
same model with isotropy. Similar studies also exist for 
eight-state $\uq{3|1}$-invariant models 
\cite{GZZ:cond-mat9709,GGZZ:cond-mat9709}.

\section{The quantum algebra $\cU_q(sl(2|1))$\label{sect:sl12}}
\subsection{Definitions}
The superalgebra ${\cal U}_{q}(sl(2|1))$ in the distinguished basis 
is the associative 
superalgebra over $\CC$ with generators 
$k_{i}^{\pm 1}$, 
$e_{i}$, $f_{i}$, ($i=1,2$) and relations
\begin{eqnarray}
  && k_1 k_2 = k_2 k_1 \;, \nonumber\\
  && k_i e_j k_i^{-1} = q^{a_{ji}} e_j\;,   \qquad\qquad
     k_i f_j k_i^{-1} = q^{-a_{ji}} f_j\;,   \label{eq:kifj}  \nonumber\\
  && e_1 f_1 - f_1 e_1 = \frac{k_1-k_1^{-1}}{q-q^{-1}}\;,   \qquad\qquad 
     e_2 f_2 + f_2 e_2 = \frac{k_2-k_2^{-1}}{q-q^{-1}}\;,
  \label{eq:e2f2}  \nonumber\\ 
  && [e_1,f_2] = 0\;,  \qquad\qquad
     [e_2,f_1] = 0\;,  \label{eq:e2f1} \nonumber\\
  && e_2^2 = f_2^2 = 0  \;, \label{eq:f2}  \nonumber\\
  && e_1^2 e_2 - (q+q^{-1}) e_1 e_2 e_1+ e_2 e_1^2 = 0  \;,
  \label{eq:serre1}  \nonumber\\ 
  && f_1^2 f_2 - (q+q^{-1}) f_1 f_2 f_1+ f_2 f_1^2 = 0  \;.
  \label{eq:serre2}   
\end{eqnarray}
The matrix $(a_{ij})$ is the distinguished Cartan matrix of $sl(2|1)$,
i.e.  
\begin{equation}
  (a_{ij}) = 
  \left( 
    \begin{array}{cc} 2 & -1 \\ 
                     -1 & 0 
    \end{array}
  \right)
  \label{eq:Cart_mat}
\end{equation}

The $\ZZ_{2}$-grading in ${\cal U}_{q}(sl(2|1))$ is uniquely defined by the 
requirement that the only odd generators are $e_{2}$ and $f_{2}$, i.e.
\begin{eqnarray}
  &&\deg\;(k_{i})=\deg\;(k_{i}^{-1})=0 \;, \nonumber\\
  &&\deg\;(e_{1})=\deg(f_{1})=0 \;, \nonumber\\
  &&\deg\;(e_{2})=\deg(f_{2})=1 \;. 
  \label{eq:degree}
\end{eqnarray}

We define a Hopf algebra structure on $\uq{2|1}$ by 
\begin{eqnarray}
  &&\Delta(k_i) = k_i \otimes k_i \;,\nonumber \\
  &&\Delta(e_i) = e_i \otimes 1
  + k_i \otimes e_i \;,\nonumber \\
  &&\Delta(f_i) = f_i \otimes k_i^{-1} 
  + 1 \otimes f_i \;,
  \label{eq:Deltadist}
\end{eqnarray}

\subsection{Centre and Scasimirs}
In the enveloping algebra ${\cal U}_{q}(sl(2|1))$, we define for
$p\in\ZZ$ the elements 
\begin{eqnarray}
  &{\cal \cQ^{(+)}}_{p} = k_{1}^{2p-1}k_{2}^{4p-2}
  & \bigg. 
  %%  (q-q^{-1})^2 %\lambda^{2} 
  \bigg\lbrace  [h_{1}+h_{2}+1][h_{2}]
  - f_{1} e_{1}
  - f_{2} e_{2} [h_{1}+h_{2}+1]
  - f_{3} e_{3} [h_{2}-1]
  \nonumber\\
  && \bigg.  
  + q^{-1} f_{3} e_{2} e_{1} k_{2} 
  + q f_{1} f_{2} e_{3} k_{2}^{-1} 
  + \left( 1 + q^{2-4p} \right) f_{2} f_{3} e_{3} e_{2} \bigg\rbrace \;,
  \label{eq:Q+}
\end{eqnarray}
and
\begin{eqnarray}
  &{\cal \cQ^{(-)}}_{p} = k_{1}^{2p-1}k_{2}^{4p-2} q^{-2p}
  & \bigg. 
  %%  (q-q^{-1})^2 %\lambda^{2} 
  \bigg\lbrace  
  q f_{2} e_{2} [h_{1}+h_{2}]
  + q f_{3} e_{3} [h_{2}-2]
  \nonumber\\
  && \bigg.  
  - q^{-1} f_{3} e_{2} e_{1} k_{2} 
  - q^3 f_{1} f_{2} e_{3} k_{2}^{-1} 
  - \left( 1 + q^{2} \right) f_{2} f_{3} e_{3} e_{2} \bigg\rbrace \;,
  \label{eq:Q-}
\end{eqnarray}
where
\begin{equation}
  e_{3}=e_{1}\;e_{2}-q^{-1}e_{2}\;e_{1} \qquad \mbox{and} \qquad
  f_{3}=f_{2}\;f_{1}-q\;f_{1}\;f_{2} \;.
  \label{eq:e3etf3}
\end{equation}
The operators $\cQ^{(\pm)}$ satisfy the following set of relations
\begin{eqnarray}
  \cQ^{(+)}_{p_1} \cQ^{(-)}_{p_2}  = 
  \cQ^{(-)}_{p_1} \cQ^{(+)}_{p_2}  = 0  && 
  \forall p_1,p_2 \in \ZZ \;,
  \label{eq:relationQ+-}
  \\
  \cQ^{(+)}_{p_1} \cQ^{(+)}_{p_2}  = \cQ^{(+)}_{p_3} \cQ^{(+)}_{p_4}  
  &\qmbox{if}& p_1+p_2=p_3+p_4 \;,
  \label{eq:relationQ++}
  \\
  \cQ^{(-)}_{p_1} \cQ^{(-)}_{p_2}  = \cQ^{(-)}_{p_3} \cQ^{(-)}_{p_4}  
  &\qmbox{if}& p_1+p_2=p_3+p_4 \;.
  \label{eq:relationQ--}
\end{eqnarray}
In the enveloping algebra $\uq{2|1}$, there are two abelian
subalgebras $\cA^{(+)}$ and $\cA^{(-)}$, generated respectively by the
operators $\cQ^{(+)}$ and $\cQ^{(-)}$. They are such that 
\begin{equation}
  \label{eq:A+A-}
  \forall x^+ \in \cA^{(+)}, \quad   \forall x^- \in \cA^{(-)}, \qquad
  x^+ x^- = x^- x^+ = 0 \;.
\end{equation}

The elements $\cQ^{(\pm)}$ allow us to build generators of the centre of 
${\cal U}_{q}(sl(2|1))$, and also a set of Scasimirs: 
if we define, for $p\in \ZZ$
\begin{eqnarray}
  \cC_p &=& \cQ^{(+)}_p + \cQ^{(-)}_p \;, \\
  \cS_p &=& \cQ^{(+)}_p - \cQ^{(-)}_p  \;,
  \label{eq:cas&scas}
\end{eqnarray}
then 
\begin{itemize}
\item $1$ and the $\cC_p$, for $p\in\ZZ$, generate the centre of
  ${\cal U}_{q}(sl(2|1))$, (for $q$ not a root of unity):
  \begin{equation}
    x \cC_p  - \cC_p x = 0 \qquad \forall x \in \uq{2|1} \;.
    \label{eq:centre}
  \end{equation}
  (See \cite{ACF,AAB}, and \cite{ABP} for the non quantised case).
\item The $\cS_p$ commute with the bosonic elements of $\uq{2|1}$ and
  anticommute with the fermionic ones (although they are themselves
  bosonic) 
  \begin{equation}
    \cS_p x  - (-1)^{\deg (x)} x \cS_p = 0 \qquad 
    \left(
      \forall x \in \uq{2|1} \quad \mbox{with homogeneous degree}
    \right) \;.
    \label{eq:scentre}
  \end{equation}
\end{itemize}
Furthermore, the $\cC_p$, $\cS_p$ obey the set of relations 
\begin{eqnarray}
  {\cal C}_{p_1} {\cal C}_{p_2}  &=& {\cal C}_{p_3} {\cal C}_{p_4} \qquad
  \hbox{if}
  \qquad p_1 + p_2 = p_3 + p_4 \;,
  \label{eq:relCCCC}
  \\
  {\cal C}_{p_1} {\cal C}_{p_2}  &=& {\cal S}_{p_3} {\cal S}_{p_4} \qquad
  \hbox{if}
  \qquad p_1 + p_2 = p_3 + p_4 \;, 
  \label{eq:relCCSS}
  \\
  {\cal C}_{p_1} {\cal S}_{p_2}  &=& {\cal S}_{p_3} {\cal C}_{p_4} \qquad
  \hbox{if}
  \qquad p_1 + p_2 = p_3 + p_4 \;, 
  \label{eq:relCSSC}
\end{eqnarray}
which is equivalent to the set 
(\ref{eq:relationQ+-},\ref{eq:relationQ++},\ref{eq:relationQ--}).
Relation (\ref{eq:relCCCC}) was given in 
\cite{ABP} for the non quantised case and in \cite{ACF} in the
quantised case.

In particular, on representations on which $\cC_p$ are different from
$0$, the quotient $\frac{\cS_p}{\cC_p}$ plays the role of $(-1)^F$,
i.e. 
\begin{eqnarray}
  \left(\frac{\cS_p}{\cC_p}\right)^2 &=& 1 \;, \\
  \frac{\cS_p}{\cC_p}  x 
  - (-1)^{\deg (x)} x \frac{\cS_p}{\cC_p} &=& 0 \qquad 
  \left(
    \forall x \in \uq{2|1} \quad \mbox{with homogeneous degree}
  \right) \;.
  \label{eq:-1^F}
\end{eqnarray}

Most Hamiltonians in the following will be constructed using
$(\rho\otimes\rho)\Delta(\cC_p)$, with then 
\begin{eqnarray}
  &{\cal C}_{p} = k_{1}^{2p-1}k_{2}^{4p-2}
  & \bigg. 
  \bigg\lbrace  [h_{1}+h_{2}+1][h_{2}]
  - f_{1}e_{1}+f_{2}e_{2}([h_{1}+h_{2}]q^{1-2p}-[h_{1}+h_{2}+1])
  \nonumber\\
  && \bigg.  
  + f_{3} e_{3} ([h_{2}-2]q^{1-2p} - [h_{2}-1])
  + (q-q^{-1}) %\lambda 
  q^{-1-p} [p] f_{3} e_{2} e_{1} k_{2} 
  \nonumber\\
  && \bigg.   
  + (q-q^{-1})  %\lambda 
  q^{2-p} f_{1} f_{2} e_{3} k_{2}^{-1}[p-1] 
  + (q-q^{-1})^2 %\lambda^{2} 
  q^{1-2p} [p] [p-1] f_{2} f_{3} e_{3} e_{2} \bigg\rbrace \;.
  \nonumber \\
  \label{eq:cas_sl21}
\end{eqnarray}

\subsection{Four dimensional representation}
We use the one-parameter four-dimensional representation, acting on
the vector space $V$ of dimension $4$, and defined (in the
distinguished basis) by
\begin{eqnarray}
  \rho(e_1) & = & -\omega q E_{23} \nonumber\\
  \rho(e_2) & = & (\lambda - \lambda^{-1}) E_{12} 
  + (q\lambda - q^{-1}\lambda^{-1}) E_{34} 
  \nonumber\\
  \rho(f_1) & = & - q^{-1} E_{32} \nonumber\\
  \rho(f_2) & = & E_{21} + E_{43} \nonumber\\
  \rho(k_1) & = & \lambda^{-1}\left(E_{11}+E_{22}
  +q^{-1}E_{33}+q^{-1}E_{44}\right)\nonumber\\
  \rho(k_2) & = & \omega\lambda^{-2}\left(E_{11}+q^{-1}E_{22}
  +q^{-1}E_{33}+q^{-2}E_{44}\right) \;,
  \label{eq:representation}
\end{eqnarray}
where $\omega=\pm 1$ is a discrete parameter that allows two different
(inequivalent) representations for each value of the continuous
parameter $\lambda\equiv q^\mu$ \cite{AAB}. The discrete parameter
$\omega$ is a remnant of the quantisation of the value of $k_1$ on the
highest weight vector in finite dimension. 

The $E_{ij}$ are the standard elementary matrices
of End($V$) given by 
\begin{equation}
  \left(E_{ij}\right)_{kl} = \delta_{ik}\delta_{jl} \;.
  \label{eq:Eij}
\end{equation}
The operators are represented by ordinary matrices, with complex
(commuting) elements. 
We do not consider supermatrices. The traces are
not supertraces.
Tensor products of representations are non-graded tensor products. 
We indeed use, as in \cite{ACF}, the non-graded coproduct defined from
the usual graded one as (in Sweedler's notation)
\begin{equation}
  \Delta^{n.g.}(x) = \sum x_{(1)} g^{\deg(x_{(2)})} \otimes x_{(2)}
  \qmbox{where}
  \Delta(x) = \sum x_{(1)} \otimes x_{(2)}
  \;, 
  \label{eq:non-grad-Delta}
\end{equation}
$g$ being the diagonal element in End($V$) defined by 
$g=\sum_{j=1}^{{\rm dim}V} (-1)^{\deg(j)} E_{jj}$ 
with $\deg(1)=\deg(4)=0$ and $\deg(2)=\deg(3)=1$. 
This is nothing but a Jordan--Wigner transformation. 
Practically, on tensor products of representations, this amounts to
the use of the graded coproduct $\Delta$, the evaluation of the
representations 
$\rho_1\otimes \rho_2$ and then application of the transformation
\begin{equation}
  E_{ij}\otimes E_{kl} \longrightarrow   
  (-1)^{\deg(j)(\deg(k)+\deg(l))} E_{ij}\otimes E_{kl} \;.
  \label{eq:tranfs-EijEkl}
\end{equation}
In the following, this will be implicitly included in the
construction. This use of ordinary matrices and non graded coproduct
is actually equivalent to the standard procedure, and leads to the
same conclusions. It is however sometimes simpler in actual computations.

The transformation from $\Delta$ to the non-graded  $\Delta^{n.g.}$ 
was used by Majid to
bosonise super Hopf algebras \cite{Majidiv}. 
It is a simple case of transmutation.  
A transformation was also defined in \cite{DGLZ:ijmpA10} and applied
to the $\cR$-matrix, which
allowed to consider non-graded Yang--Baxter equations.

\section{Braid group representation}

Explicit computation show that
\begin{eqnarray}
  (\rho\otimes\rho)\Delta(\cC_p) = - q^{-1}\lambda^{8p-4}
  \left(
    [2\mu][2\mu+1]\cO_0 + q^{2p-1} [2\mu][2\mu+2]\cO_1 \right.
    \nonumber\\
    \left.
      + q^{4p-2} [2\mu+1][2\mu+2]\cO_2 \right) \;,
  \label{eq:deltaCp}
\end{eqnarray}
where the expression of the operators $\cO_a$ is given later in Equation
(\ref{eq:O}).

The operators $\cO_a$ satisfy the relations
\begin{eqnarray}
  \cO_a \cO_b &=& \delta_{a,b} \cO_a \nonumber\\
  \cO_0 + \cO_1 + \cO_2 &=& \Id \;.
  \label{eq:proj}
\end{eqnarray}

The operators $\cO_0$, $\cO_1$, $\cO_2$ are actually 
projectors on the representations of dimension $4$, $8$
and $4$, respectively, that enter in the decomposition of the tensor
product $\rho\otimes\rho$ (using $\Delta$). 

Inverting (\ref{eq:deltaCp}) allows to express these projectors
directly in terms of evaluations on the tensor product
$\rho\otimes\rho$ of some Casimir operators $\cC_p$:
\begin{eqnarray}
  \cO_0 &=& \frac{ q^4 \lambda^{-8p-4}}{[2\mu][2\mu+1](q^4-1)(q^2-1)}
  (\rho\otimes\rho) \Delta
  \left( 
    - q^{3} \lambda^8 \cC_p + (q + q^{-1})\cC_{p+1} 
    -  q^{-3} \lambda^{-8} \cC_{p+2} 
  \right) \;, \nonumber \\ 
  \cO_1 &=& \frac{ q^{-2p+4} \lambda^{-8p-4}}
  {[2\mu][2\mu+2](q^4-q^2)(q^2-1)}
  (\rho\otimes\rho) \Delta
  \left( 
     q^2 \lambda^8 \cC_p - (q^2 + q^{-2})\cC_{p+1} 
     +  q^{-2}\lambda^{-8} \cC_{p+2} 
  \right) \;, \nonumber \\ 
  \cO_2 &=& \frac{ q^{-4p+4} \lambda^{-8p-4}}
  {[2\mu+1][2\mu+2](q^4-q^2)(q^4-1)}
  (\rho\otimes\rho) \Delta
  \left( 
    - q \lambda^8 \cC_p + (q + q^{-1})\cC_{p+1} - 
    q^{-1} \lambda^{-8} \cC_{p+2} 
  \right) \;, \nonumber\\
  \label{eq:O-fct-de-Cp}
\end{eqnarray}
where, again, $\lambda = q^\mu$.

As a consequence of (\ref{eq:proj}), the algebra generated by all the
$\uq{2|1}$ 
invariant operators $(\rho\otimes\rho)\Delta(\cC_p)$ is 
$\Vect(\cO_0,\cO_1,\cO_2)=\Vect(\Id,\cO_0,\cO_2)\subset 
\mbox{End}(V\otimes V)$.
Within this algebra, we look for operators $b$ 
satisfying the braid group relations 
\begin{eqnarray}
  b_i b_{i\pm 1} b_i & = & b_{i\pm 1} b_i b_{i\pm 1} \;, \\
  b_i b_j & = & b_j b_i \qmbox{for} |i-j|\ge 2 \;,
  \label{eq:tresses}
\end{eqnarray}
where 
\begin{equation}
  b_i \equiv b_{i,i+1} = 1\otimes\cdots\otimes b\otimes \cdots\otimes
  1 \;, 
  \label{eq:bii+1}
\end{equation}
in which $b$ occupies positions $i,i+1$.

We find two non trivial solutions to these equations, given by 
\begin{equation}
  b = -q \Id + q\lambda \frac{[2\mu]}{[\mu]} \cO_0
  + \lambda^{-1} \frac{[2\mu+2]}{[\mu+1]} \cO_2 \;,
  \label{eq:solbraid1}
\end{equation}
the other one being its inverse $b^{-1}$.
\begin{equation}
  b^{-1} = -q^{-1} \Id + q^{-1}\lambda^{-1} \frac{[2\mu]}{[\mu]} \cO_0
  + \lambda \frac{[2\mu+2]}{[\mu+1]} \cO_2 \;.
  \label{eq:solbraid2}
\end{equation}
These are the only solutions for generic $\lambda=q^\mu$. For particular
values of $\lambda$, i.e. $\lambda=\pm q^{-1/2}$ for instance, there
are other solutions to the braid relations, which can lead to
Temperley--Lieb algebra \cite{FoeLinRod:mplA12}.

We define $x=(\lambda-\lambda^{-1}) (q \lambda-q^{-1} \lambda^{-1}) $,
and $y=([\mu][\mu+1])^{1/2}=x^{1/2}/(q-q^{-1})$ 
including the freedom for a sign in $y$. 

The explicit expressions for $b$ and $b^{-1}$ are 
\begin{eqnarray}
  b&=&
     q \lambda^2  E_{11}\otimes E_{11}   
  + ( q \lambda^2 - q ) E_{11}\otimes E_{22}   
  + ( q \lambda^2 - q ) E_{11}\otimes E_{33}   
  +  x E_{11}\otimes E_{44} 
  \nonumber\\ &&
  + q \lambda  (E_{12}\otimes E_{21}+E_{21}\otimes E_{12})   
  +  q^{-1/2} x^{1/2}  (E_{12}\otimes E_{43}+E_{21}\otimes E_{34})   
  \nonumber\\ &&
  +  q \lambda  \omega (E_{13}\otimes E_{31}+E_{31}\otimes E_{13})    
  - q^{1/2} x^{1/2} \omega (E_{13}\otimes E_{42}+E_{31}\otimes E_{24})    
  \nonumber\\ &&
  + q  \omega (E_{14}\otimes E_{41}+E_{41}\otimes E_{14})    
  - q  E_{22}\otimes E_{22}   
  + ( q^{-1} - q ) E_{22}\otimes E_{33}   
  \nonumber\\ &&
  + ( q^{-1} \lambda^{-2} - q ) E_{22}\otimes E_{44}   
  - \omega (E_{23}\otimes E_{32}+E_{32}\otimes E_{23})    
  \nonumber\\ &&
  + \lambda^{-1} \omega (E_{24}\otimes E_{42}+E_{42}\otimes E_{24})    
  - q  E_{33}\otimes E_{33}   
  + ( q^{-1} \lambda^{-2} - q ) E_{33}\otimes E_{44}   
  \nonumber\\ &&
  + \lambda^{-1} (E_{34}\otimes E_{43}+E_{43}\otimes E_{34})   
  +  q^{-1} \lambda^{-2} E_{44}\otimes E_{44}   \;,
  \label{eq:b}
\end{eqnarray}

\begin{eqnarray}
  b^{-1}&=&
   q^{-1} \lambda^{-2}  E_{11}\otimes E_{11}   
  + q^{-1} \lambda^{-1} (E_{12}\otimes E_{21}+E_{21}\otimes E_{12})   
  \nonumber\\ &&
  + q^{-1} \lambda^{-1} \omega (E_{13}\otimes E_{31}+E_{31}\otimes E_{13})    
  + q^{-1} \omega (E_{14}\otimes E_{41}+E_{41}\otimes E_{14})    
  \nonumber\\ &&
  + ( q^{-1} \lambda^{-2} - q^{-1} ) E_{22}\otimes E_{11}   
  - q^{-1} E_{22}\otimes E_{22}   
  - \omega (E_{23}\otimes E_{32}+E_{32}\otimes E_{23})   
  \nonumber\\ &&
  - q^{-1/2} x^{1/2} \omega (E_{24}\otimes E_{31}+E_{42}\otimes E_{13})    
  + \lambda \omega (E_{24}\otimes E_{42}+E_{42}\otimes E_{24})    
  \nonumber\\ &&
  + ( q^{-1} \lambda^{-2} - q^{-1} ) E_{33}\otimes E_{11}   
  + ( q - q^{-1} ) E_{33}\otimes E_{22}   
  - q^{-1} E_{33}\otimes E_{33}   
  \nonumber\\ &&
  + q^{1/2} x^{1/2} (E_{34}\otimes E_{21}+E_{43}\otimes E_{12})   
  + \lambda (E_{34}\otimes E_{43}+E_{43}\otimes E_{34})   
  \nonumber\\ &&
  + x E_{44}\otimes E_{11} 
  + ( q \lambda^2 - q^{-1} ) E_{44}\otimes E_{22}   
  + ( q \lambda^2 - q^{-1} ) E_{44}\otimes E_{33}   
  \nonumber\\ &&
  +  q \lambda^2  E_{44}\otimes E_{44} \;.
  \label{eq:b-1}
\end{eqnarray}

\section{A cubic algebra, Baxterisation and exact
   solvability\label{sect:baxterisation}} 

These solutions satisfy the cubic equations
\begin{eqnarray}
  \left(b_i + q \right)
  \left(b_i - q \lambda^2\right)
  \left(b_i - q^{-1} \lambda^{-2} \right) &=& 0 \;, \\
  \left(b_i^{-1} + q^{-1} \right)
  \left(b_i^{-1} - q \lambda^2\right)
  \left(b_i^{-1} - q^{-1} \lambda^{-2} \right) &=& 0 \;. 
  \label{eq:trilinear}
\end{eqnarray}

The explicit expressions for the projectors $\cO_a$ can be obtained
from (\ref{eq:b},\ref{eq:b-1}) by inverting
(\ref{eq:solbraid1},\ref{eq:solbraid2}), i.e.,
\begin{eqnarray}
  \cO_0 &=& \frac{[\mu]}{[2\mu][2\mu+1]}
  \left(
    [\mu+1] \Id + \frac{1}{q-q^{-1}}\left(\lambda b -
    \lambda^{-1}b^{-1}\right) 
  \right) \;,
  \nonumber\\
  \cO_1 &=& \frac{[\mu][\mu+1]}{[2\mu][2\mu+2]}
  \left(\left(q\lambda^2+q^{-1}\lambda^{-2}\right) \Id 
    - b - b^{-1}
  \right) \;,
  \nonumber\\
  \cO_2 &=& \frac{[\mu+1]}{[2\mu+1][2\mu+2]}
  \left(
    [\mu] \Id + \frac{1}{q-q^{-1}}\left( - q^{-1}\lambda^{-1} b +
    q\lambda b^{-1}\right) 
  \right) \;.
  \label{eq:O}
\end{eqnarray}

We can use the cubic equations (\ref{eq:trilinear}) in a 
Baxterisation procedure \cite{Jones} 
to get solution of the Yang--Baxter algebra
\begin{eqnarray}
  && \vR_{i,i+1}(u) \vR_{i+1,i+2}(u+v) \vR_{i,i+1}(v) =   
  \vR_{i+1,i+2}(v) \vR_{i,i+1}(u+v)
  \vR_{i+1,i+2}(u)  \;, \nonumber\\
  && \vR_{i,i+1}(u) \vR_{j,j+1}(v)  = 
  \vR_{j,j+1}(v) \vR_{i,i+1}(u) \qmbox{for} |i-j|\ge 2 \;.
  \label{eq:YBA}
\end{eqnarray}
The matrix $\vR$ is related to the matrix  $\cR$ by $\vR=\cP\cR$,
the operator 
$\cP$ being the permutation map $\cP:x\otimes y\mapsto y\otimes x$.

In the simplest case where $b_i$ satisfies a quadratic relation (Hecke
case), it is possible to find a linear
combination of $b$ and $b^{-1}$ that is solution of the Yang--Baxter
algebra (Baxterisation).

We look here for solutions of the Yang--Baxter algebra (\ref{eq:YBA})
with $\vR(u)$ in the linear span of $\Id$, $b$, $b^{-1}$ with
coefficients depending on $u$.

We find the solution
\begin{equation}
  \vR_{i,i+1}(u) = 1 + \frac{1}{x} 
  \left( 
    (e^u-1) b_{i} + (e^{-u}-1) b_{i}^{-1}) 
  \right) \;,
  \label{eq:R}
\end{equation}
relying on the fact that $b$ obeys the supplementary relation
\begin{eqnarray}
 0 &=& b_i b_{i\pm 1}^{-1} b_i 
  - b_{i\pm 1} b_i^{-1} b_{i\pm 1} 
  - b_i^{-1} b_{i\pm 1} b_i^{-1} 
  + b_{i\pm 1}^{-1} b_i b_{i\pm 1}^{-1} \nonumber\\
  &&- x (b_i b_{i\pm 1}^{-1} - b_i^{-1} b_{i\pm 1} 
  - b_{i\pm 1} b_i^{-1}+b_{i\pm 1}^{-1} b_i)
  \nonumber\\
  && -  x (q^{-1}   (b_i-b_{i\pm 1}) -  q  (b_i^{-1}-b_{i\pm 1}^{-1}))
  \label{eq:relsuppl}
\end{eqnarray}
or equivalently
\begin{eqnarray}
  (b_i-x) b_{i\pm 1}^{-1} (b_i -x)
  &-& b_i^{-1} (b_{i\pm 1}-x) b_i^{-1} 
  \nonumber\\
  = \quad
  (b_{i\pm 1}-x) b_i^{-1} (b_{i\pm 1} -x)
  &-& b_{i\pm 1}^{-1} (b_i-x) b_{i\pm 1}^{-1} \;.
  \label{eq:relsuppl2}
\end{eqnarray}

The algebra satisfied by the operators $b_i$ is then given by
(\ref{eq:algebra1}-\ref{eq:algebra4}). 
It  is sufficient to define an exactly solvable
periodic spin chain. This algebra was already used in
\cite{GeWuXue:1991} to obtain solutions of the Yang--Baxter algebra
(\ref{eq:YBA}).

We notice that we do not have a full BWM algebra: in
the algebra generated by $b_i$, $b_i^{-1}$, the operators $e_i$ 
such that
\begin{equation}
  \label{eq:e1^2}
  e_i ^2 = \alpha e_i
\end{equation}
satisfy neither 
\begin{equation}
  \label{eq:toBMW_TL}
  e_i e_{i\pm 1} e_i = \alpha' e_i 
\end{equation}
nor
\begin{equation}
  \label{eq:toBMW+}
  e_i b_{i\pm 1} e_i = \alpha'' e_i \;.
\end{equation}

The relations
(\ref{eq:algebra1}--\ref{eq:algebra4}) are 
nevertheless enough to ensure that the $\vR$-matrix (\ref{eq:R})
satisfies the Yang--Baxter algebra. 

The $\vR$-matrix with spectral parameter $u$ 
satisfies the inversion relation:
\begin{equation}
  \vR(u)\vR(-u) = \zeta(u) \;,
  \label{eq:inv-rel}
\end{equation}
with 
\begin{equation}
  \zeta(u) = e^{-2u}(e^u-\lambda^{-2})(e^u-\lambda^2)
  (e^u-q^2\lambda^{-2})(e^u-q^{-2}\lambda^2)/x^2 \;.
  \label{eq:zeta}
\end{equation}
It has PT symmetry:
\begin{equation}
  \cR_{21}(u) \equiv \cP \cR_{12}(u) \cP = \cR_{12}(u)^{t_1t_2} \;.
  \label{eq:PT-sym}
\end{equation}
It satisfies also the crossing unitarity property 
\cite{ReshSemTS:lpm19,MezNep:jpa24}: 
\begin{equation}
  %%  \cR_{12}(u)^{t_1} M_1 \cR_{21}(-u-2\rho)^{t_1} M_1^{-1} = \xi(u) 
  \cR_{12}(u)^{t_1} M_1 \cR_{21}(-u-2\rho)^{t_1} M_1^{-1} = \xi(u+\rho) \;,
    \label{eq:cross-unit}
\end{equation}
with 
\begin{equation}
  \rho = \ln q \;, \qquad M = \matdiag{1}{-1}{-q^2}{q^2}
  \label{eq:M}
\end{equation}
and 
\begin{equation}
  \xi(u) =
  %%  - (e^u-1)(1-e^{-u})(q^2e^u-1)(1-q^{-2}e^{-u})/x^2
  - (q^{-1}e^u-1)(1-qe^{-u})(qe^u-1)(1-q^{-1}e^{-u})/x^2 \;.
  \label{eq:xi}
\end{equation}

We define the row-to-row transfer matrix on a closed chain 
as $Tr_0 \cT(u)$, where
$\cT(u)$ is the monodromy matrix given by
\begin{equation}
  \cT(u) = \cR_{0L}(u)\cR_{0\ L-1}(u)\cdots \cR_{01}(u) \;.
  \label{eq:Rchain}
\end{equation}
The Yang--Baxter algebra satisfied by $\cR$ ensures that transfer
matrices with different spectral parameters commute, i.e.
\begin{equation}
  [Tr_0 \cT(u), Tr_0 \cT(v)]=0 \qquad \forall u,v \;.
  \label{eq:commT}
\end{equation}

{}From the $\cR$-matrix one can extract a spin chain 
Hamiltonian with nearest
neighbour interaction
\begin{equation}
  \cH_{\rm per} = x
  \left.\frac{d}{du}
  \right|_{u=0} \cT(u) 
  = \sum_{i=1}^{L-1} \cH_{i\  i+1}   + \cH_{L\, 1} \;,
  \label{eq:Hinteg}
\end{equation}
with
\begin{equation}
  \cH_{i,i+1} = x
  \left.\frac{d}{du}\right|_{u=0} \vR_{i,i+1}(u) = b_i - b_i^{-1} \;.
  \label{eq:Hintegi}
\end{equation}
With periodic boundary conditions, this Hamiltonian also commutes with
all the transfer matrices, which is the requirement for its exact
solvability.
The Hamiltonian with ordinary periodic boundary conditions is however not
$\uq{2|1}$-invariant. A method was developed in
\cite{FoersterKarowi} to construct a periodic Hamiltonian which is
still $\uq{2|1}$-invariant, by adding a ``$\cH_{L1}$''-type term which
is not completely local. A simpler solution is also presented in
\cite{FoersterLinks:jpa30}.

\section{Two site quantum chain Hamiltonian\label{sect:twositeHam}}

To obtain a model of interacting electrons, 
we will use, as in \cite{GHLZ:pla212}  
the following interpretation of the states of the representation
in terms of fermionic states: 
\begin{equation}
  \left| 1 \right\rangle = \left| \uparrow\downarrow \right\rangle 
  = c^\dagger_{\downarrow} c^\dagger_{\uparrow}
  \left| \emptyset \right\rangle \qquad
  \left| 2 \right\rangle = \left| \downarrow \right\rangle 
  = c^\dagger_{\downarrow}  \left| \emptyset \right\rangle \qquad
  \left| 3 \right\rangle = \left| \uparrow \right\rangle 
  = c^\dagger_{\uparrow} \left| \emptyset \right\rangle \qquad
  \left| 4 \right\rangle = \left| \emptyset \right\rangle \;.
  \label{eq:fermions}
\end{equation}
We will also use
\begin{eqnarray}
  n_{\uparrow} &=& c^\dagger_{\uparrow} c_{\uparrow} = E_{11}+E_{33}
  \;, \\
  n_{\downarrow} &=& c^\dagger_{\downarrow} c_{\downarrow} =
  E_{11}+E_{22} \;, \\
  n &=& n_{\uparrow} + n_{\downarrow} = 2E_{11} + E_{22} + E_{33} \;.\\
  \label{eq:n_up}
\end{eqnarray}

The expression of the spin chain Hamiltonian obtained in this case is
given by
\begin{equation}
  \cH^{dist} = \cH_{hop} + \cH_{diag}^{dist} \;,
  \label{eq:Hdist}
\end{equation}
where
\begin{eqnarray}
  \cH_{hop} &=& \left( 
  c^\dagger_{\uparrow i+1}c^\dagger_{\downarrow i+1} 
  c_{\downarrow i}c_{\uparrow i}
  +
  c^\dagger_{\uparrow i}c^\dagger_{\downarrow i}
  c_{\downarrow i+1}c_{\uparrow i+1} 
  \right)\nonumber\\
  &+& 
  \left( c^\dagger_{\uparrow i+1} c_{\uparrow i} 
    + c^\dagger_{\uparrow i} c_{\uparrow i+1} 
  \right)
  \left\{ -[\mu] + n_{\downarrow i}\left([\mu]+ q^{-1/2}y\right)
    + n_{\downarrow i+1}\left([\mu]- q^{1/2}y\right)
  \right.\nonumber\\
  &&
  \left. \qquad
    + n_{\downarrow i}n_{\downarrow i+1}\left(-[\mu]+[\mu+1]
    + (q^{1/2}-q^{-1/2})y\right)
  \right\} \nonumber\\
  &+& \omega
  \left( c^\dagger_{\downarrow i+1} c_{\downarrow i} 
    + c^\dagger_{\downarrow i} c_{\downarrow i+1} 
  \right)
  \left\{ -[\mu] + n_{\uparrow i}\left([\mu]- q^{1/2}y\right)
    + n_{\uparrow i+1}\left([\mu]+ q^{-1/2}y\right)
  \right.\nonumber\\
  &&
  \left. \qquad
    + n_{\uparrow i}n_{\uparrow i+1}\left(-[\mu]+[\mu+1]
    + (q^{1/2}-q^{-1/2})y\right)
  \right\} 
  \label{eq:Hhop}
\end{eqnarray}
and
\begin{eqnarray}
  \cH_{diag}^{dist} &=& 
  n_{\uparrow i}n_{\downarrow i}
  + n_{\uparrow i+1}n_{\downarrow i+1}
  - [2\mu+1] \nonumber\\
  &+& q^{\mu+1}[\mu] (n_{\uparrow i} + n_{\downarrow i})
  + q^{-\mu-1}[\mu] (n_{\uparrow i+1} + n_{\downarrow i+1}) \;,
  \label{eq:Hdiagdist}
\end{eqnarray}
where $\mu$ is related to the parameter of the representation
$\lambda$ by $\lambda=q^\mu$. By construction, the creation and
annihilation operators on different sites commute. A Jordan--Wigner
transformation can restore the standard anticommutation property.

This exactly solvable Hamiltonian with two parameters $\lambda=q^\mu$
and $q$ was already considered  in \cite{BarKluZit,GHLZ:pla212}.
In \cite{GHLZ:pla212}, it was obtained as the derivative of the
spectral parameter 
$\vR$-matrix of the four dimensional representation of
$\uq{sl{2|1}}$. The eigenstates of the periodic model are found in
\cite{HibGouLin:jpa29} using the algebraic Bethe ansatz.

\section{Reflection equations and open chain Hamiltonian\label{sect:refleq}}
\subsection{Reflection equations}

We can also get an exactly solvable and $\uq{2|1}$-invariant 
{\em open} chain Hamiltonian by solving the reflection equations 
\cite{Cherednik:1984,Sklyanin:1988,KulSkl:1991,MezNep:jpa24,%
MezNep:1991ijmp}
\begin{equation}
  \cR_{12}(u-v) \cK_1^-(u) \cR_{21}(u+v) \cK_2^-(v) 
  =
  \cK_2^-(v)  \cR_{12}(u+v) \cK_1^-(u) \cR_{21}(u-v) 
  \label{eq:RE-}
\end{equation}
and
\begin{eqnarray}
  && 
  \cR_{12}(-u+v) \cK_1^+(u)^{t_1} M_1^{-1} 
  \cR_{21}(-u-v-2\rho) M_1 \cK_2^+(v)^{t_2} 
  =\nonumber\\
  && \qquad\qquad\qquad\qquad
  \cK_2^+(v)^{t_2} M_1 \cR_{12}(-u-v-2\rho) 
  M_1^{-1} \cK_1^+(u)^{t_1} \cR_{21}(-u+v) \;.
  \qquad\qquad
  \label{eq:RE+}
\end{eqnarray}

The simplest solution for these equations is \cite{MezNep:1991ijmp}
\begin{equation}
  \cK^-(u)=\Id   \qmbox{and}   \cK^+(u)=M \;.
  \label{eq:K-+sol0}
\end{equation}
This is
always a solution when the spectral parameter $\cR$-matrix is obtained
via self-Baxterisation \cite{Jones}, 
i.e. when the $\vR$-matrix belongs to the
algebra generated by $b_i$, since in this case $\vR$-matrices with different
spectral parameters commute:
\begin{equation}
  \left[
    \vR(u),\vR(v)
  \right] = 0 
  \qquad \forall u,v \in \CC \;.
  \label{eq:RuRv}
\end{equation}
The matrix $M$ may in this case be interpreted as a Markov trace, as
in \cite{FoersterKarowii}.

More generally, there are two 
diagonal one parameter solutions for $\cK^-(u)$ (up to an
overall function of $u$), given by
\begin{eqnarray}
  &&
  \cK_a^-(u) = \frac{1}{(1+C)(1+q^2C)} \cdot \nonumber\\
  &&
  \cdot \matdiag{(e^{-u}+C)(e^{-u}+q^2C)\hspace{-.4cm}}
  {\hspace{-.4cm} (e^{u}+C)(e^{-u}+q^2C) \hspace{-.4cm}}
  {\hspace{-.4cm} (e^{u}+C)(e^{-u}+q^2C) \hspace{-.4cm}}
  {\hspace{-.4cm} (e^{u}+C)(e^{u}+q^2C)}
  \nonumber\\
  \label{eq:K-sol1}
\end{eqnarray}
and
\begin{equation}
  \cK_b^-(u) =
  \frac{1}{1+C}\matdiag{e^{-u}+C}{e^{-u}+C}{e^{u}+C}{e^{u}+C} \;.
  \label{eq:K-sol2}
\end{equation}
Solutions for $\cK^+(u)$ are given by \cite{MezNep:1991ijmp}
\begin{equation}
  K^+(u) = K^-(-u-\rho)^t M \;.
  \label{eq:K+sol}
\end{equation}
Note that the number of one parameter diagonal solutions is the same
as for the supersymmetric $t$--$J$ model \cite{Gonzalez-Ruiz:npb424}
and is equal to the rank of the underlying algebra.

\subsection{Open chain transfer matrix and exactly solvable Hamiltonian}

Using the Reflection Equations (\ref{eq:RE-}), (\ref{eq:RE+}), 
and the Yang--Baxter algebra (\ref{eq:YBA}), one can
prove that the  double-row transfer matrices $t(u)$
\cite{MezNep:jpa24} 
\begin{eqnarray}
  t(u) &=& \zeta(u)^{-L} \mbox{tr}\ \cK^+(u) T(u) \cK^-(u) T(-u)^{-1} \\
  &=& \mbox{tr}_0 \cK_0^+(u)
  \vR_{L0}(u)\vR_{L-1,L}(u)\cdots\vR_{23}(u)\vR_{12}(u) \cdot
  \nonumber\\  && \ \ \ 
  \cdot \ \cK_1^-(u) 
  \vR_{12}(u)\cdots\vR_{23}(u)\cdots\vR_{L-1,L}(u)\vR_{L0}(u) 
  \label{eq:double-row}
\end{eqnarray}
commute for different values of $u$
\cite{Sklyanin:1988,KulSkl:1991,MezNep:1991ijmp,GouldLinks:1996}.

We then compute
\begin{equation}
  \left. 
    \frac{dt(u)}{du}
  \right|_{u=0} - 
  \left.
    \frac{d}{du}\mbox{tr}_0\cK_0^+(u)
  \right|_{u=0} = 
  \left( 
    \mbox{tr}_0 \cK_0^+(0)
  \right)
  \left(
    2\sum_{j=1}^{L-1} \cH_{j,j+1} +
    \left.
      \frac{d}{du}\cK_1^-(u)
    \right|_{u=0}
  \right)
  + 2 \mbox{tr}_0 \cK_0^+(0) \cH_{L0} \;.
  \label{eq:dt/du}
\end{equation}
It is standard to use this expression, divided by 
$\mbox{tr}_0 \cK_0^+(0)$, 
to get a spin chain Hamiltonian with nearest neighbour interaction. By
construction, this Hamiltonian commutes with $t(u)$ for all values of
$u$ and it is hence exactly solvable \cite{Sklyanin:1988}.

This operation however provides nothing here, since, for all the diagonal
solutions for $\cK^+$, we have $\mbox{tr}_0 \cK_0^+(0) = 0$.
This phenomenon was
noticed in \cite{GouldLinks:1996}, and explained by the use of typical
representations, which implies $\mbox{tr}M=0$ (actually
$\mbox{Str}M=0$ if no bosonisation is performed).
A method was found there to prove
that, in the case 
\begin{equation}
  \cK^-(u)=1 \qmbox{and} \cK^+(u)=M \;,
  \label{eq:K=1}
\end{equation}
the quantum chain Hamiltonian 
\begin{equation}
  \sum_{j=1}^{L-1} \cH_{j,j+1}
  \label{eq:Honly}
\end{equation}
still commuted with $t(u)$ for all values of $u$. 
The $\uq{2|1}$ symmetry is built-in in this case, since 
the expression of the Hamiltonian 
(\ref{eq:Honly}) contains only the coproduct of some Casimir operators
(See equations (\ref{eq:Hintegi}), (\ref{eq:solbraid1}),
(\ref{eq:solbraid2}) and (\ref{eq:O-fct-de-Cp}) 
which provide the expression of $\cH_{i,i+1}$ in terms of some
$(\rho\otimes\rho)\Delta(\cC_p)$). 
This Hamiltonian 
is then both exactly solvable and quantum group invariant.

Another way to obtain a Hamiltonian with local interaction in the
cases when $\mbox{tr}_0 \cK_0^+(0) = 0$ is to 
take the second derivative of $t(u)$ at $u=0$. 
This method was also
used in \cite{CuernoGonz:1993}, where the vanishing of the factor was
due to the fact that $q$ was such that $q^4=1$. 
It applies also with the solutions for $\cK^+$ different from $M$ and
given by (\ref{eq:K+sol}) and (\ref{eq:K-sol1}) or (\ref{eq:K-sol2}).
\begin{eqnarray}
  \left. 
    \frac{d^2t(u)}{du^2}
  \right|_{u=0} 
  &= &
  \left( 
    \left.
      2 \frac{d}{du}\mbox{tr}_0 \cK_0^+(u)
    \right|_{u=0}
    + 4 \mbox{tr}_0 
    \left( 
      \cK_0^+(0) \cH_{L0}
    \right)
  \right) \times \qquad\qquad\qquad\qquad
  \nonumber\\
  &&\qquad \qquad \qquad\qquad \times
  \left(
    2\sum_{j=1}^{L-1} \cH_{j,j+1} +
    \left.
      \frac{d}{du}\cK_1^-(u)
    \right|_{u=0}
  \right)
  \nonumber\\
  && +  A_1 + A_2 + A_3 + A_4 \;,
  \label{eq:d2t/du2}
\end{eqnarray}
with
\begin{eqnarray}
  A_1 &=&   
  \left.
    \frac{d^2}{du^2}\mbox{tr}_0\cK_0^+(u)
  \right|_{u=0} \;,
  \label{eq:A1}
  \\
  A_2 &=& 4 \mbox{tr}_0 
  \left( 
    \left.
      \frac{d}{du} \cK_0^+(u)
    \right|_{u=0}
    \cH_{L0}
  \right) \;,
  \label{eq:A2}
  \\
  A_3 &=& 2 \mbox{tr}_0 \cK_0^+(0) 
  \left.
    \frac{d^2}{du^2} \vR_{L0}(u)
  \right|_{u=0} \;,
  \label{eq:A3}
  \\
  A_4 &=& 2 \mbox{tr}_0 
  \left( 
    \cK_0^+(0) \cH_{L0} \cH_{L0}
  \right) \;.
  \label{eq:A4}
\end{eqnarray}

Now the factor 
$  \left( 
  \left. 2 \frac{d}{du}\mbox{tr}_0 \cK_0^+(u) \right|_{u=0}
  + 4 \mbox{tr}_0 
  \left(  \cK_0^+(0) \cH_{L0}\right)
\right)  = 2
\left.  \frac{d}{du} \mbox{tr}_0 \cK_0^+(u) \vR_{L0}^2 \right|_{u=0}$
in front of the Hamiltonian of interest can be chosen to be non-zero. 
Moreover, it is proportional to the
identity, so that we can use  
\begin{equation}
  \frac{1}{  4
    \left.  \frac{d}{du} \mbox{tr}_0 \cK_0^+(u) \vR_{L0}^2
    \right|_{u=0}
    }
  \left.
    \frac{d^2t(u)}{du^2}
  \right|_{u=0}
  \label{eq:d2tdu2/norm}
\end{equation}
as a spin chain Hamiltonian with nearest neighbour interaction.

The term 
$ \left. \frac{d}{du}\cK_1^-(u) \right|_{u=0} $
contributes to a boundary term on site $1$.

The term $A_1$ obviously contributes only as constant. 
The terms $A_2$, $A_3$ and $A_4$ contribute to boundary
terms on the last site $L$ of the chain. 
Note that the sum $A_1 + A_2 + A_3 + A_4$ is equal to 
\begin{equation}
  A_1+A_2+A_3+A_4 = 
  \left.  \frac{d^2}{du^2} \mbox{tr}_0 \cK_0^+(u) \vR_{L0}^2
  \right|_{u=0} \;.
  \label{eq:A1A2A3A4}
\end{equation}

The expression of the exactly solvable Hamiltonian with open boundary
condition  is then
\begin{equation}
  \label{eq:Ham-open}
  \cH_{open} = \sum_{j=1}^{L-1} \cH_{j,j+1} + \frac12
    \left.
      \frac{d}{du}\cK_1^-(u)
    \right|_{u=0}
    + 
    \frac
    {\left.  \frac{d^2}{du^2} \mbox{tr}_0 \cK_0^+(u) \vR_{L0}^2
      \right|_{u=0}}
    {4\left.  \frac{d}{du} \mbox{tr}_0 \cK_0^+(u) \vR_{L0}^2
    \right|_{u=0}} \;.
\end{equation}
{}From the expressions of the boundary terms in (\ref{eq:Ham-open}),
one can prove that, if the solution of the reflections equations are
multiplied by arbitrary functions of $u$, the Hamiltonian is left
unchanged (up to constant terms).

\subsection{Integrable boundary terms}
We use the construction of section \ref{sect:twositeHam} for the
expression of the bulk term $\cH_{j,j+1}=\cH_{j,j+1}^{dist}$ of
Eq. (\ref{eq:Hdist}), (\ref{eq:Hhop}) and (\ref{eq:Hdiagdist}).
We then include the results of section \ref{sect:refleq} for the
boundary terms (inserting the $\vR$ matrix of section
\ref{sect:baxterisation}). 
We get 
\begin{equation}
  \label{eq:Hdist-open}
  \cH_{open}^{dist} = \sum_{j=1}^{L-1} \cH_{j,j+1}^{dist} 
  + \cB_1 + \cB_L \;.
\end{equation}

The boundary term 
$ \cB_1 = \left. \frac{d}{du}\cK_1^-(u) \right|_{u=0}$
on site $1$ takes one of the forms
\begin{eqnarray}
  && \cB_{1}^0 = 0 \qmbox{(in the case $\cK^- = 1$)} 
  \\
  & \mbox{or} & \nonumber
  \\
  && \cB_{1}^a = 
  \frac{-1}{(1+C_{-})(1+q^2C_{-})}
  \Big\{
  (2 + C_{-} + q^2C_{-}) E_{11} +  (1 + C_{-}) \left(E_{22} + E_{33}\right) 
  \Big\}
  \label{eq:bord1a}
  \\
  & \mbox{or} & \nonumber
  \\
  &&   \cB_{1}^b = 
  \frac{-1}{(1+C_{-})} \left(E_{11} + E_{22}\right) \;,
  \label{eq:bord1b}
\end{eqnarray}
(mutually exclusive) depending on the choice of the solution ($\cK_a^-$
or $\cK_b^-$) for the
matrix $\cK^-$. It depends on the parameter $C_{-}\equiv C$ from
(\ref{eq:K-sol1}) or (\ref{eq:K-sol2}). 

These expressions read, in terms of number of particles
\begin{eqnarray}
  && \cB_{1}^0 = 0 \;,
  \\
  && \cB_{1}^a = 
  \frac{-1}{(1+C_{-})(1+q^2C_{-})}
  \Big\{
  (q^2-1) C_{-} n_{\uparrow 1} n_{\downarrow 1} 
  +  (1 + C_{-}) \left(n_{\uparrow 1} + n_{\downarrow 1}\right) 
  \Big\} \;,
  \label{eq:bord1a_n}
  \\
  &&   \cB_{1}^b = 
  \frac{-1}{(1+C_{-})} n_{\downarrow 1} \;.
  \label{eq:bord1b_n}
\end{eqnarray}

The boundary term 
$ \cB_{L} =   \frac
    {\displaystyle\left.  \frac{d^2}{du^2} \mbox{tr}_0 \cK_0^+(u) \vR_{L0}^2
      \right|_{u=0}}
    {\displaystyle 4\left.  \frac{d}{du} \mbox{tr}_0 \cK_0^+(u) \vR_{L0}^2
    \right|_{u=0}}
$
on site $L$ takes one of the forms 
\begin{eqnarray}
  && \cB_{L}^0 = 0 \qmbox{(in the case $\cK^+ = M$)} 
  \\
  & \mbox{or} & \nonumber
  \\
  && \cB_{L}^a = 
  %%  \frac{1}{(1+C_{+})(1+q^2C_{+})}
  %%  \Big\{
  %%  (2 + C_{+} + q^2C_{+}) E_{11} +  (1 + C_{+}) 
  %%  \left(E_{22} + E_{33}\right) 
  %%  \Big\}
  \frac{1}{(1+q^{-1}\lambda^{-2}C_{+})(1+q\lambda^{-2}C_{+})} \cdot
  \nonumber\\
  && \qquad \cdot
  \Big\{
  (2 + q^{-1}\lambda^{-2}C_{+}  q^{-1}\lambda^{-2}C_{+}) E_{11} +  
  (1 + q \lambda^{-2}C_{+}) \left(E_{22} + E_{33}\right) 
  \Big\}
  \label{eq:bordLa}
  \\
  & \mbox{or} & \nonumber
  \\
  && \cB_{L}^b = 
  \frac{1}{(1+q^{-1}\lambda^{-2}C_{+})} \left(E_{11} + E_{22}\right) \;,
  \label{eq:bordLb}
\end{eqnarray}
depending on the choice of
solution for the matrix $\cK^+$ (which is independent of the choice
for $\cK^-$). It depends on a parameter $C_{+}$ coming from
(\ref{eq:K-sol1}) or (\ref{eq:K-sol2}) when used as solutions for $\cK^+$ 
given by (\ref{eq:K+sol}).

These expressions read, in terms of number of particles and after a
redefinition of the parameter $C_{+}$ that eliminates the dependence
in $\lambda$,
\begin{eqnarray}
  && \cB_{L}^0 = 0 \;,
  \\
  && \cB_{L}^a = 
  \frac{1}{(1+C'_{+})(1+q^2C'_{+})}
  \Big\{
  (1-q^2) C'_{+} n_{\uparrow L} n_{\downarrow L} 
  +  (1 + q^2 C'_{+}) \left(n_{\uparrow L} + n_{\downarrow L}\right) 
  \Big\} \;,
  \label{eq:bordLa_n}
  \\
  && 
  \cB_{L}^b = 
  \frac{1}{(1+C'_{+})} n_{\downarrow L} \;.
  \label{eq:bordLb_n}
\end{eqnarray}

As we will see in the next section, there exists a non trivial choice
for the boundary terms $\cB_{1}^b$ and $\cB_{L}^b$ that leads to an
exactly solvable Hamiltonian {\em with $\uq{2|1}$ invariance}.

\section{Another spin chain Hamiltonian: using the fermionic basis of 
$\uq{2|1}$ \label{sect:Hamil-ferm}}

Alternatively, we could have used form the beginning 
the fermionic basis to describe the
quantum algebra. In this basis, the Cartan matrix is 
\begin{equation}
  (a_{ij})_{ferm} = 
  \left( 
    \begin{array}{cc} 0 & -1 \\ 
                     -1 & 0 
    \end{array}
  \right)
  \label{eq:Cart_mat_ferm}
\end{equation}
The generators $K_1$, $K_2$, $E_1$, $E_2$, $F_1$, $F_2$ in the
fermionic basis are, in terms of the generators 
in the distinguished basis
\begin{eqnarray}
  K_1 = k_1^{-1}k_2^{-1} &&
  K_2 = k_2 \nonumber\\
  E_1 = e_3 &&
  E_2 = f_2  k_2^{-1} \nonumber\\
  F_1 = -f_3 &&
  F_2 = k_2  e_2 \;.
  \label{eq:dist2ferm}
\end{eqnarray}
As algebras, $\uq{2|1}$ in both bases are identical. Only the choices
of simple root are different. However, the Hopf structure are not
identical: the coproduct in the fermionic basis is given by
\begin{eqnarray}
  &&\tilde\Delta(K_i) = K_i \otimes K_i \;,\nonumber \\
  &&\tilde\Delta(E_i) = E_i \otimes 1
  + K_i \otimes E_i \;,\nonumber \\
  &&\tilde\Delta(F_i) = F_i \otimes K_i^{-1} 
  + 1 \otimes F_i \;,
  \label{eq:Deltaferm}
\end{eqnarray}
which, in terms of the distinguished generators, is different from
(\ref{eq:Deltadist}) (See \cite{ACF}), and will produce (using the
same algorithm as for the distinguished case) a quantum
chain Hamiltonian different from (\ref{eq:Hdist}):
\begin{equation}
  \cH^{ferm} = \cH_{hop} + \cH_{diag}^{ferm}
  \label{eq:Hferm}
\end{equation}
with
\begin{eqnarray}
  \cH_{diag}^{ferm} &=& 
  n_{\uparrow i}n_{\downarrow i}
  + n_{\uparrow i+1}n_{\downarrow i+1}
  - [2\mu+1] 
  \nonumber\\
  &+& q^{\mu+1}[\mu] (n_{\uparrow i} + n_{\downarrow i+1})
  + q^{-\mu-1}[\mu] (n_{\uparrow i+1} + n_{\downarrow i}) \;.
  \label{eq:Hdiagferm}
\end{eqnarray}
The Hamiltonians obtained with the distinguished basis and with the
fermionic basis are actually very close to each other: the only
difference is in boundary terms, which are symmetric in $\uparrow$ and
$\downarrow$  in the distinguished case, but not in the
fermionic one.
When summed over the chain, the difference of the Hamiltonians
$\cH_{open}^{ferm}$ and $\cH_{open}^{dist}$ (without integrable
boundary terms added) is indeed 
\begin{equation}
  \cH_{open}^{ferm} - \cH_{open}^{dist} = 
  \sum_{j=1}^{L-1} 
  \left( \cH_{diag\ j,j+1}^{ferm} - \cH_{diag\ j,j+1}^{dist} 
  \right)
  = \frac{x}{q-q^{-1}} \big( n_{\downarrow L} - n_{\downarrow 1} \big)
  \;.
  \label{eq:Hdiff}
\end{equation}
The Hamiltonian $\cH_{open}^{ferm}$ is actually {\em equal} to the
Hamiltonian (\ref{eq:Hdist-open}) obtained with the distinguished
basis, now including the integrable boundary terms $\cB_{1}^b$
(\ref{eq:bord1b_n}) and $\cB_{L}^b$ (\ref{eq:bordLb_n})
coming from the {\em second} 
solution (\ref{eq:K-sol2}) of the reflection equations
(\ref{eq:RE-}), (\ref{eq:RE+}), for the
particular choice of parameters 
\begin{equation}
  C_{-} = \frac{q-q^{-1}}{x} -1 \qquad 
  C_{+} = q\lambda^2 C'_{+} = q \lambda^2 \left(\frac{q-q^{-1}}{x} 
  -1 \right) \;.
  \label{eq:C+-sol}
\end{equation}

Although different, the two Hopf structures defined by 
(\ref{eq:Deltadist}) and (\ref{eq:Deltaferm})
are equivalent
\cite{KhoTolb} through a twist \cite{Resh:lmp20}
\begin{equation}
  \tilde\Delta(a) = \cF \Delta(a) \cF^{-1} \;,
  \label{eq:Resh-twist}
\end{equation}
satisfying
\begin{eqnarray}
  && (\epsilon  \otimes 1) \cF = (1 \otimes \epsilon) \cF = 1 \;, \\
  && (\cF \otimes 1) (\Delta \otimes 1) \cF = 
  (1\otimes \cF) (1\otimes \Delta) \cF \;.
  \label{eq:F-prop}
\end{eqnarray}
It was indeed proved in \cite{KhoTolb} that an operator $\cF$
satisfying (\ref{eq:Resh-twist}) could be obtained as the
factor of the universal $\cR$-matrix of $\uq{2|1}$ related with the
fermionic root which defines the super-Weyl reflection that relates
the two bases. 

This implies that open quantum chains built with the two-site 
Hamiltonians (\ref{eq:Hdist}) and
(\ref{eq:Hferm}) are equivalent, the equivalence matrix being given by
\begin{equation}
  \Big(\rho\otimes \cdots \otimes\rho \Big) \cF^{(L)} \;,
  \label{eq:rhoF}
\end{equation}
$\cF^{(L)}$ being defined recursively as 
\begin{equation}
   \cF^{(L)} \equiv (\cF \otimes 1^{\otimes^{L-1}}) 
   (\Delta \otimes 1^{\otimes^{L-1}}) \cF^{(L-1)} \;.
  \label{eq:FL}
\end{equation}

As in \cite{ACF}, this equivalence is simple
for the two site Hamiltonians (i.e. for (\ref{eq:Hdist}) and
(\ref{eq:Hferm}) themselves). However, it becomes highly non trivial
for longer chains, the reason being that the equivalence produced by
the twist is non local. 

In \cite{FoeLinRod:1997b}, twists of the coproduct are applied to the
supersymmetric $t$--$J$ model and to the supersymmetric Hubbard model
with pair hopping (\ref{eq:Hdist}). This leads to multiparametric 
Hamiltonians. The effects of these twists are visible in the bulk term
of the Hamiltonian, in contrast with the action of our twist which
relates the distinguished construction to the fermionic one, and which
affects only boundary terms.

\section{Another example}

We can also obtain $\uq2 \otimes U(1)$ invariant Hamitonians as 
\begin{equation}
  \cH = \sum_{i=1}^{L-1} 1\otimes \cdots \otimes 
  \underbrace{(\rho\otimes\rho)\Delta( \mbox{Pol} \{\cQ^{(+)}_p
  ,\cQ^{(-)}_p \} )}_{ 
    \mbox{sites\ } i,i+1}
  \otimes \cdots \otimes 1 \;.
\end{equation}
Choosing the four dimensional representation with the fixed parameter
$\lambda=q^{-1/2}$, and taking a polynomial in $\cQ^{(+)}$ only, 
we get for instance 
\begin{eqnarray}
  \cH_{i,i+1}^{TL} &=& 
  c^\dagger_{\uparrow i+1}c^\dagger_{\downarrow i+1} 
  c_{\downarrow i}c_{\uparrow i}
  +
  c^\dagger_{\uparrow i}c^\dagger_{\downarrow i}
  c_{\downarrow i+1}c_{\uparrow i+1} 
  - S^+_i S^-_{i+1}
  - S^-_i S^+_{i+1}
  \nonumber\\
  &+& 
  \left(  c^\dagger_{\uparrow i+1} c_{\uparrow i} 
    - c^\dagger_{\uparrow i} c_{\uparrow i+1} 
  \right) \omega
  \left\{ q^{-1} n_{\downarrow i} + q n_{\downarrow i+1} - 
    (q+q^{-1}) n_{\downarrow i} n_{\downarrow i+1}
  \right\} \nonumber\\
  &+& 
  \left( - c^\dagger_{\downarrow i+1} c_{\downarrow i} 
    + c^\dagger_{\downarrow i} c_{\downarrow i+1} 
  \right)
  \left\{ n_{\uparrow i} + n_{\uparrow i+1} - 
    2 n_{\uparrow i} n_{\uparrow i+1}
  \right\} \nonumber\\
  &+& 
  ( n_{\uparrow i} - n_{\uparrow i+1} )
  \left\{ q^{-1} n_{\downarrow i} - q n_{\downarrow i+1} - 
    (q-q^{-1}) n_{\downarrow i} n_{\downarrow i+1}
  \right\} 
  \label{eq:H_TL}
\end{eqnarray}
which satisfies the 
Temperley--Lieb algebra
\begin{eqnarray}
  && b_i^2 = 0
  \\
  && b_i b_{i\pm 1} b_i  =  b_i \\
  && b_i b_j  =  b_j b_i \qmbox{for} |i-j|\ge 2 \;.
  \label{eq:TLA}
\end{eqnarray}
Such  Hamiltonians were found in 
\cite{Links:jpa29,FoeLinRod:mplA12}. 
It was noticed that, although not Hermitian, they led to Hermitian
Hamiltonian when multiplied by 
$(1-2n_{\downarrow i} -2n_{\uparrow i}+4n_{\uparrow i}n_{\downarrow i})$ 
(the parity operator on one site), the result satisfying also a
Temperley--Lieb algebra (with non vanishing square). 
 
It could also be of interest to investigate the use of the Hamiltonian
(\ref{eq:H_TL}) itself for reaction-diffusion processes \cite{ADHR}.

\bigskip
\paragraph{Acknowledgments:} I warmfully thank P.~Pearce and
V.~Rittenberg for interesting discussions, and for pointing out some
references. 
I am  also indebted to Prof. I. Musson who convinced me 
about the existence of Scasimirs in the case of $sl(m|n)$.
I also thank M.~Bauer and V.~Lafforgues for fruitful discussions. 

\appendix

\section{Appendix: Scasimirs of $\cU(sl(2|1))$}

We give in this appendix the expressions of the Scasimirs of
non-deformed superalgebra $\cU(sl(2|1))$.

The Scasimir of $osp(2|1)$ appeared in 
\cite{PaisRitt,Pinc,Lesniewski}. In  \cite{Lesniewski}, the expression
of the Scasimir is also given in the $q$-deformed case. 

The proof of existence of Scasimir operators for $osp(1|2n)$ was given
in \cite{Musson:1997ja,ABFscasimirs}, where it was also proved that
the Scasimir was the square root of a Casimir element of degree $2n$. 
An explicit expression of the Scasimir is written in \cite{ABFscasimirs}. 

The existence of Scasimir operators in the case of $sl(m|n)$ is known
to Musson \cite{Musson:private}.

The classical superalgebra $sl(2|1)$ is defined by the relations
\begin{eqnarray}
  && [h_1, h_2 ] = 0 \;, \nonumber\\
  && [h_i, e_j ] = a_{ji} e_j \;,   \qquad\qquad
     [h_i, f_j ] = -a_{ji} f_j \;,   \label{eq:hifj}  \nonumber\\
  && [e_1, f_1 ] = h_1 \;,   \qquad\qquad 
     [e_2, f_2 ]_+ = h_2 \;,  \label{eq:e2f2cl}  \nonumber\\
  && [e_1,f_2] = 
     [e_2,f_1] = 0\;,  \label{eq:e2f1cl} \nonumber\\
  && [e_2,e_2]_+ = [f_2,f_2]_+ = 0  \;, \label{eq:f2cl}  \nonumber\\
  && [e_1,e_3] = [f_1, f_3] = 0 \label{eq:e1e3cl} \;,
\end{eqnarray}
where 
\begin{equation}
  e_{3} = [e_{1}, e_{2}] \qquad \mbox{and} \qquad
  f_{3} = [f_{2}, f_{1}] \;.
  \label{eq:e3f3cl}
\end{equation}
The last relations in 
(\ref{eq:e1e3cl}) may also be written as Serre relations
\begin{eqnarray}
  && e_1^2 e_2 - 2 e_1 e_2 e_1+ e_2 e_1^2 = 0  \;,
  \label{eq:serre1cl}  \nonumber\\ 
  && f_1^2 f_2 - 2 f_1 f_2 f_1+ f_2 f_1^2 = 0  \;.
  \label{eq:serre2cl}   
\end{eqnarray}

We define the elements $\cQ^{(\pm)}_p$ 
of the non-quantum $\cU(sl(2|1))$  as 
\begin{eqnarray}
  &{\cal \cQ^{(+)}}_{p} & = \bigg\lbrace 
  h_2 (h_1+h_2+1) - f_1 e_1 - f_2 e_2 (h_1+h_2+1) - f_3 e_3 (h_2-1)
  \nonumber\\
  &&  + f_1 f_2 e_3 + f_3 e_2 e_1 + f_2 f_3 e_3 e_2 \bigg\rbrace
  (- h_1 - 2h_2 - 1)^{p-2} 
  \nonumber\\
  && + f_2 f_3 e_3 e_2  (- h_1 - 2h_2 + 1)^{p-2} 
  \label{eq:Q+class}
\end{eqnarray}
and
\begin{eqnarray}
  &{\cal \cQ^{(-)}}_{p} & = \bigg\lbrace 
  f_2 e_2 (h_1+h_2) + f_3 e_3 (h_2-2)
  \nonumber\\
  &&  - f_1 f_2 e_3 - f_3 e_2 e_1 - 2 f_2 f_3 e_3 e_2 \bigg\rbrace
  (- h_1 - 2h_2 )^{p-2} 
  \label{eq:Q-class}
\end{eqnarray}
for $p\ge 2$. Their sum $\cC_p$ and difference $\cS_p$ are, 
respectively, Casimir
operators and Scasimirs of $\cU(sl(2|1))$, i.e. they satisfy the
classical analogues of (\ref{eq:centre},\ref{eq:scentre}). The relations 
(\ref{eq:relationQ+-},\ref{eq:relationQ++},\ref{eq:relationQ--},
\ref{eq:relCCCC},\ref{eq:relCCSS},\ref{eq:relCSSC}) are still valid as
long as the indices $p_i$ are greater or equal to 2.
Notice that the classical operators $\cQ^{(\pm)}_p$, $\cC_p$ and
$\cS_p$ are not 
the limits as $q$ goes to $1$ of the corresponding quantum ones, but
rather limits of some linear combinations of them (See \cite{ACF}). 

Discussions with M. Bauer and V. Lafforgue led to an expression of 
$\cS_2$ in terms of antisymmetrised products of fermionic operators 
$e_i$, $f_i$, $i=2,3$ only, as for $osp(1|2n)$ in
\cite{ABFscasimirs}. 
This seems to be possible for more general superalgebras. 

%% \bibliography{bibref,publ}

\begingroup\raggedright\endgroup

\end{document}